\documentclass[
    superscriptaddress,
    aip,
	 jap,
    preprintnumbers,
    twocolumn,
    reprint,
    nobalancelastpage,
    nonbibnotes,
    fleqn,
    a4paper,
	 floatfix
]{revtex4-1}
\newcommand{\twocolumnmode}{true}

\usepackage{graphicx}
\usepackage{amsmath}
\usepackage{amsfonts}
\usepackage{ifthen}
\usepackage[cp1250]{inputenc}
\usepackage{color}
\usepackage[
    colorlinks,
    linkcolor=blue,
    anchorcolor=blue,
    citecolor=blue,
    urlcolor=blue,
    menucolor=blue,
    filecolor=blue
]{hyperref}

\newcommand{\emtext}[1]{{\em #1\/}}

\newcommand{\Li}[1]{\mathop{\rm Li_{#1}}\nolimits}

\begin{document}

\title{Patterning of dielectric nanoparticles using dielectrophoretic forces generated by ferroelectric polydomain films}

\author{P. \surname{Mokr\'{y}}}
\email{pavel.mokry@tul.cz} %

\affiliation{Institute of Mechatronics and Computer Engineering, Technical University of Liberec, CZ--46117 Liberec, Czech Republic}

\author{M. \surname{Marvan}}
\affiliation{Faculty of Mathematics and Physics, Charles University, 18000 Praha 8, Czech Republic}

\author{J. \surname{Fousek${}^{1}$}}
\noaffiliation

\date{\today}

\begin{abstract}
A theoretical study of a dielectrophoretic force, i.e. the force acting on an 
electrically neutral particle in the inhomogeneous electric field, which is 
produced by a ferroelectric domain pattern, is presented. It has been shown by 
several researchers that artificially prepared domain patterns with given 
geometry in ferroelectric single crystals represent an easy and flexible 
method for patterning dielectric nanoobjects using dielectrophoretic forces. 
The source of the dielectrophoretic force is a strong and highly inhomogeneous 
(stray) electric field, which exists in the vicinity of the ferroelectric 
domain walls at the surface of the ferroelectric film. We analyzed 
dielectrophoretic forces in the model of a ferroelectric film of a given 
thickness with a lamellar 180${}^\circ$ domain pattern. The analytical formula 
for the spatial distribution of the stray field in the ionic liquid above the 
top surface of the film is calculated including the effect of free charge 
screening. The spatial distribution of the dielectrophoretic force produced by 
the domain pattern is presented. The numerical simulations indicate that the 
intersection of the ferroelectric domain wall and the surface of the 
ferroelectric film represents a trap for dielectric nanoparticles in the case 
of so called positive dielectrophoresis. The effects of electrical neutrality 
of dielectric nanoparticles, free charge screening due to the ionic nature of 
the liquid, domain pattern geometry, and the Brownian motion on the mechanism 
of nanoparticle deposition and the stability of the deposited pattern are 
discussed.
\end{abstract}



\maketitle


\section{Introduction}

A well-known fact follows from the Maxwell theory that the inhomogeneous 
electromagnetic field induces a force even on electrically neutral particles. 
However, for a long time there has been no attention on how this phenomenon 
could be practically applied. A new impetus in this field was provided in a 
1951 paper by Pohl \cite{Pohl.JApplPhys.22.1951} where he studied the 
influence of inhomogeneous electrostatic field on the motion of small plastic 
particles suspended in insulating dielectric fluids. Pohl named this 
phenomenon \emtext{dielectrophoresis} (DEP) and it was shown in his 1978 
monograph \cite{Pohl.1978} that DEP forces can be usefully used for the 
manipulation of small particles. This book, in fact, motivated studies of 
other possibilities to use the electric field for trapping or controlling the 
motion of small particles.

At the same time, in 1980s, a new field of \emtext{nanotechnology} emerged and 
DEP has become an important and intensively studied tool for the manipulation 
with nanoparticles. Examples to be mentioned here are the DEP assembly of 
nanowires from nanoparticle suspensions \cite{Hermanson.Science.294.2001}, DEP 
precise positioning of carbon nanotubes 
\cite{Banerjee.JVacSciTechnolB.24.2006}, particle separation using DEP 
\cite{Ramos.JPhysD.31.1998,Morgan.BiophysJ.77.1999,
Gascoyne.Electrophor.23.2002}, DEP formation of nanodroplets 
\cite{Jones.JApplPhys.89.2001} or nanofibers \cite{Lukas.JApplPhys.103.2008}, 
etc. In all the aforementioned applications, the inhomogeneous electric field, 
which is essential for the existence of DEP forces, is achieved by properly 
arranged electrodes. The crucial point related to nanotechnology is that, when 
there is a wish to work with nanoparticles, it is necessary to deposit 
very thin electrodes which are very close to each other. This task can be 
obviously managed \cite{Hughes.Nanotechnol.11.2000} but, in fact, it is 
usually technologically demanding. In addition, a fixed set of electrodes 
unfortunately does not allow flexible changes in the nanoparticle pattern 
geometry, which may be necessary in some advanced applications.

In order to overcome disadvantages of the fixed electrode approach, attention 
of researchers has been drawn to answer the question: would it be more useful 
to use an inhomogeneous field existing above the surface of a ferroelectric 
film with a domain pattern? The interest of scientists in this question has 
been stimulated by a great advance that has recently been achieved in a 
precise writing of ferroelectric domains of nanometer dimensions. Several 
methods for nanoscale domain control in ferroelectric films have been 
developed by several groups of researchers. Classical examples of domain 
engineering to be mentioned here are the domain poling with patterned electrodes
\cite{Matsumoto.ElLett.27.1991,Yamada.ApplPhysLett.62.1993,Grilli.ApplPhysLett.87.2005}, 
domain patterning by scanning probe microscopy
\cite{Kolosov.PhysRevLett.74.1995,Gruverman.AnnuRevMatRes.28.1998,Liu.JpnJApplPhys.44.2005,Rodriguez.ApplPhysLett.86.2005,Cho.ApplPhysLett.81.2002},
and focused electron or ion beam domain engineering
\cite{Son.JCrystGrowth.281.2005,Li.JpnJApplPhys.44.2005,Li.JMatRes.21.2006}. 
Details about advanced nanoscale domain engineering methods can be found in 
the recent review paper by Li and Bonnell\cite{Li.AnnRevMatRes.38.2008}. The 
aforementioned methods make it possible to prepare arbitrary ferroelectric 
domain patterns in a flexible way.

In fact, there exist two different means to use ferroelectric domains for 
nanoparticles patterning. The first approach is based on the use of 
domain-specific chemical reactions and it is called {\em ferroelectric 
lithography\/}\cite{Kalinin.AdvMat.16.2004,Li.CeramInt.34.2008}. Using this 
method, the direct assembly of virus particles has been realized by Dunn {\em 
et al.}\cite{Dunn.ApplPhysLett.85.2004}The second approach is based on the use 
of physical DEP forces that are produced by bound charges at the surface of 
the ferroelectric film with a domain pattern, which has been recently reported 
by Ke \cite{Ke.JApplPhys.101.2007} {\em et al.} and Grilli {\em et al.} 
\cite{Grilli.ApplPhysLett.92.2008}.

It is the second approach that has motivated the theoretical analysis 
presented below, where we will address the study of the DEP forces produced by 
lamellar 180${}^\circ$ domain pattern. In Sec.~\ref{sec_model}, we first 
present the details of our model of a ferroelectric film with 180${}^\circ$ 
domain domain pattern. Section~\ref{sec_potential} presents a straightforward 
calculation of the electrostatic potential and the electric field produced by 
the ferroelectric domain pattern. We will cover the general configuration of 
the domain pattern including the effect of free charge screening as well as 
two cases of a special interest. First, the domain pattern with wide domains 
compared to the thickness of the ferroelectric film 
(Sec.~\ref{subsec_wide_domains}) and the system with a dense domain pattern 
(Sec.~\ref{subsec_dense_domains}). In Sec.~\ref{sec_DEP}, calculation and 
numerical simulations of the DEP forces produced domain patterns in four 
systems with different configurations are presented and compared. 
Section~\ref{sec_discussion} presents a brief discussion of the effect of 
electrical neutrality of dielectric nanoparticles on the mechanism of 
nanopatterning (Sec.~\ref{subsec_electroneutrality}), the effect of free 
charge screening on the inhomogeneous field and the value of DEP force 
(Sec.~\ref{subsec_screening}), the role of domain pattern geometry on the 
value of the DEP force produced by the ferroelectric domain pattern 
(Sec.~\ref{subsec_geometry}), the role of the Brownian motion on the stability 
of the nanoparticle patterns (Sec.~\ref{subsec_brownian_motion}), and, finally,
 the estimation of the values of important numerical parameters that are 
essential in the design of a ferroelectric system used for patterning of 
nanoparticles (Sec.~\ref{subsec_conclusions}).


\section{Geometry and material properties of the ferroelectric film}
\label{sec_model}

%
%
\begin{figure}[t]
\begin{center}
	\includegraphics[width=85mm]{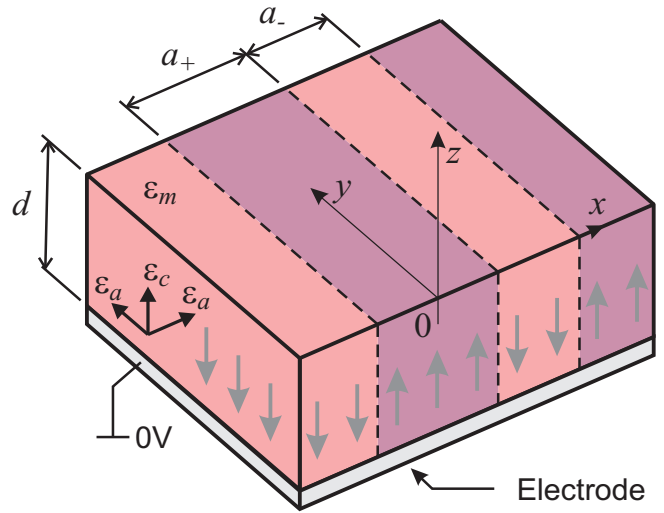}
\end{center}
\caption{
Geometry of the ferroelectric film of the thickness $d$ with the $180^\circ$ 
domain pattern. Gray arrows indicate the orientation of the vector of 
spontaneous polarization in the film. Dashed lines indicate positions of the 
180${}^\circ$ domain walls. Symbols $a_+$ and $a_-$ stand for the widths of 
domains where the vector of the spontaneous polarization is oriented along and 
against the orientation of the $z$-axis of the attached coordinate system, 
respectively. Symbols $\varepsilon_a$ and $\varepsilon_c$ stand for the 
relative permittivity tensor components of the ferroelectric, symbol 
$\varepsilon_m$ stands for the permittivity of the liquid.
}
\label{fig01}
\end{figure}
%
%
%
Figure \ref{fig01} shows the geometry of the considered model of a 
ferroelectric film with a lamellar 180${}^\circ$ ferroelectric domain pattern 
crossing the film thickness $d$. The space above the top surface of the film 
is filled with an ionic liquid of permittivity $\varepsilon_m$ and with a 
given concentration of free charge carriers. The bottom surface of the 
ferroelectric film is attached to the ground electrode. We consider that the 
top and bottom surfaces of the film are perpendicular to the ferroelectric 
axis $z$ of the attached Cartesian coordinate system. The axis $x$ of the 
coordinate system is perpendicular to the system of 180${}^\circ$ domain 
walls. Using the ``hard ferroelectric" approximation, we express the electric 
displacement of the ferroelectric as a sum of the linear dielectric response 
of the crystal lattice to the electric field and the constant spontaneous 
polarization $P_0$ (whose orientation differs from domain to domain):
\begin{subequations}
\begin{eqnarray}
    D_x &=&  \varepsilon_0 \varepsilon_a E_x, \label{eq01a} \\
    D_z &=&  \varepsilon_0 \varepsilon_c E_z \pm P_0, \label{eq01b}
\end{eqnarray}
\end{subequations}
where $\varepsilon_{c}$ and $\varepsilon_{a}$ are the components of the 
permittivity tensor of the crystal lattice in the directions parallel and 
perpendicular to the ferroelectric axis, respectively, and $\varepsilon_0$ is 
the vacuum permittivity. We consider that the vector of spontaneous 
polarization is perpendicular to the film surfaces and its magnitude $P_0$ is 
constant within each domain. For given widths $a_+$ and $a_-$ of domains, 
where the vector of the spontaneous polarization is oriented along and against 
the direction of the $z$-axis of the attached coordinate system, respectively, 
we define the average (net) spontaneous polarization of the ferroelectric 
layer $P_N=P_0\,(a_+-a_-)/(2a)$, where $a=(a_++a_-)/2$ is the domain spacing.

The abrupt change of the spontaneous polarization to zero at the interface of 
the ferroelectric film and the ionic liquid yields the appearance of a bound 
charge of the surface density $\sigma_b(x)$, which is the source of the 
spatially nonuniform electric field. The spatial distribution of the bound 
charge due to the discontinuous change of the normal component of the 
spontaneous polarization at the ferroelectric film surface in the direction of 
the $x$ axis can be written in the form:
\begin{equation}
 \sigma_b(x) = P_N + \sum_{n=1}^{\infty}
   \frac{4 P_0}{\pi n}
	\sin\left[
   	\frac{n\pi}2
      \left(
      	1 + \frac{P_N}{P_0}
      \right)
   \right]
   \cos{n k x},
	\label{eq02}
\end{equation}
where $k = \pi/a$.


\section{Electrostatic potential produced by the domain pattern} 
\label{sec_potential}

In this section we present the calculation of the electrostatic potential 
produced by the ferroelectric domain pattern. Since the dielectric properties 
of the ferroelectric film and the ionic liquid are quite different, we use the 
following symbols $\varphi$ and $\varphi_f$ for the electrostatic potential in 
the ionic liquid (for $z>0$) and the ferroelectric film (for $0>z>-d$), 
respectively. In our analysis, we consider the general situation where the 
ionic liquid above the top surface of the ferroelectric film contains free 
charge carriers of given concentration. In such a situation, the electric 
field is screened by the free charge carriers and vanishes within some typical 
distance from the surface of the ferroelectric film. In order to calculate the 
spatial distribution of the electrostatic potential in the system, we adopt 
the Debye-H\"uckel theory\cite{Debye.PhysZeit.24.1923} (DHT) and the 
aforementioned functions for the electrostatic potential should satisfy: first,
 the following partial differential equations of DHT and electrostatics
\begin{subequations}
\label{eq03}
\newlength{\tmpA}
\settowidth{\tmpA}{$\varepsilon_a\varepsilon_{a}$}
\newlength{\tmpB}
\settowidth{\tmpB}{${}_{ff}$}
\begin{eqnarray}
	\hspace{\tmpA}\hspace{\tmpB}
   \frac{\partial^2\varphi}{\partial x^2} +
   \frac{\partial^2\varphi}{\partial z^2} 
	- \frac{\varphi}{\lambda_D^2} &=& 0, \label{eq03a} \\
   \varepsilon_a\frac{\partial^2\varphi_f}{\partial x^2} +
   \varepsilon_c\frac{\partial^2\varphi_f}{\partial z^2} &=& 0, \label{eq03b} 
\end{eqnarray}
where 
\begin{equation}
	\lambda_D = \sqrt{\frac{\varepsilon_0\varepsilon_m k_B T}{2 N_A e^2 I}}
	\label{eq03c}
\end{equation}
is {\em the Debye screening length}, $k_B$ is the Boltzmann constant, $T$ is 
the thermodynamic temperature, $N_A$ is the Avogadro number, $e$ is the 
electron charge, $I = (1/2) \sum_i c_i Z_i^2$ is {\em the ionic strength} of 
the liquid, where $c_i$ and $Z_i$ are the molar concentration and the charge 
number of the $i$-th particular ion in the liquid.

Second, the boundary conditions for the continuity of the electrostatic 
potential
\newlength{\tmpCa}
\settowidth{\tmpCa}{$\varphi_f$}
\newlength{\tmpCb}
\settowidth{\tmpCb}{$0$}
\begin{eqnarray}
   \varphi &=& \varphi_f \hspace{\tmpCb}\qquad \mbox{at } z=0,\label{eq03d} \\
   \varphi_f &=& 0 \hspace{\tmpCa}\qquad \mbox{at } z=-d.\label{eq03e} 
\end{eqnarray}
Third, the boundary condition for the continuity of the tangential component 
of the electric field
\begin{equation}
	\frac{\partial\varphi}{\partial x} =
	\frac{\partial\varphi_f}{\partial x} \qquad \mbox{at } z=0,\label{eq03f} 
\end{equation}
Fourth, the boundary condition for the continuity of the normal component of 
the electric displacement
\begin{equation}
	- \varepsilon_m\frac{\partial\varphi}{\partial z}
   + \varepsilon_c\frac{\partial\varphi_f}{\partial z} 
   =
   \frac{\sigma_b(x)}{\varepsilon_0}  \qquad \mbox{at } z=0,\label{eq03g} 
\end{equation}
\end{subequations}

It is a straightforward task to show that the the following functions in the 
form of Fourier series represent the solution of the electrostatic problem 
given by Eqs.~(\ref{eq03}):
\ifthenelse{\equal{\twocolumnmode}{true}}{
%
%
%
\begin{subequations}
\label{eq04}
\begin{eqnarray}
	\varphi &=&
	\frac{d P_N}{\varepsilon_0\varepsilon_c\xi_D}e^{-\frac{\hat{z}}{k\lambda_D}} 
	+ 
	\frac{C_E}{k}
	\sum_{n=1}^{\infty}
		\sin\left(
			\frac {n\pi}2 + n\hat{p}
	   \right)
	\times
	\nonumber \\  
	\lefteqn{
		\frac{
			e^{-n\hat{z}\eta_D(n)}
		}{
			n^2
			\left[\eta_D(n)+g\coth{nR}\right]
		}\,
		\cos{(n \hat{x})},
	\label{eq04a}
	}
	\\
	\varphi_f &=& 
	\frac{P_N\left(d+\hat{z}/k\right)}{\varepsilon_0\varepsilon_c\xi_D} 
	+
	\frac{C_E}{k}
	\sum_{n=1}^{\infty}
		\sin\left(
			\frac {n\pi}2 + n\hat{p}
	   \right)
	\times
	\nonumber \\  
	\lefteqn{
		\frac{
			\sinh{(nR + nc \hat{z})}
		}{
			n^2
			\left[\eta_D(n)\sinh{nR}+g\cosh{nR}\right]	
		}\,
		\cos{(n \hat{x})},
	\label{eq04b}
	}
\end{eqnarray}
}{
%
%
%
\begin{subequations}
\label{eq04}
\begin{eqnarray}
	\varphi &=&
	\frac{d P_N}{\varepsilon_0\varepsilon_c\xi_D}e^{-\frac{\hat{z}}{k\lambda_D}}
	+ 
	\frac{C_E}{k}
	\sum_{n=1}^{\infty}
		\sin\left(
			\frac {n\pi}2 + n\hat{p}
	   \right)
		\frac{
			e^{-n\hat{z}\eta_D(n)}
		}{
			n^2
			\left[\eta_D(n)+g\coth{nR}\right]
		}\,
		\cos{(n \hat{x})},
	\label{eq04a}
	\\
	\varphi_f &=& 
	\frac{P_N\left(d+\hat{z}/k\right)}{\varepsilon_0\varepsilon_c\xi_D} 
	+
	\frac{C_E}{k}
	\sum_{n=1}^{\infty}
		\sin\left(
			\frac {n\pi}2 + n\hat{p}
	   \right)
		\frac{
			\sinh{(nR + nc \hat{z})}
		}{
			n^2
			\left[\eta_D(n)\sinh{nR}+g\cosh{nR}\right]	
		}\,
		\cos{(n \hat{x})},
	\hspace{7mm}
	\label{eq04b}
\end{eqnarray}
}
where $\hat{x}=kx$, $\hat{z}=kz$, $\hat{p}=\pi P_N/(2P_0)$, 
$c=\sqrt{\varepsilon_a/\varepsilon_c}$, $g=c\varepsilon_c/\varepsilon_m$, 
$R=ckd$, and
\begin{eqnarray}
	{C_E} &=& \frac{4P_0}{\pi\varepsilon_0\varepsilon_m}, 
	\label{eq04c} \\
	\xi_D &=& 1+\frac{\varepsilon_m d}{\varepsilon_c \lambda_D}, 
	\label{eq04d} \\
	\eta_D(n) &=& \sqrt{1+\left(nk\lambda_D\right)^{-2}}.
	\label{eq04e} 
\end{eqnarray}
\end{subequations}
In the special case of the absence of free charge carriers in the liquid, i.e. 
infinite value of the Debye screening length $\lambda_D$, the values of 
parameters $\xi_D$ and $\eta_D(n)$ for all $n\ge 1$ tend to 1 and the solution 
of the electrostatic problem given by Eqs.~(\ref{eq03}) reduces 
down to the form that is already available in literature 
\cite{Kopal.Ferroelectrics.202.1997}, however in the special case of the 
ferroelectric film in a vacuum, i.e. for $\varepsilon_m = 1$.

Equations (\ref{eq04}) represent the exact solution of the electrostatic 
potential produced by the domain pattern. For the sake of clearer graphical 
presentation, it is convenient to use the normalized value $\Phi$ of the 
electrostatic potential: $\Phi = \varphi (k/C_E)$. In the following 
subsections, we will present two special situations in the system with the 
absence of free charge carriers, i.e. $\lambda_D\gg a$, in which the above 
expressions can be further simplified.

%
%
\begin{figure*}[t]
\begin{center}
\ifthenelse{\equal{\twocolumnmode}{true}}{
%
%
%
\begin{minipage}[t]{\textwidth}
\begin{minipage}[t]{0.69\textwidth}
\begin{minipage}[t]{\textwidth}
\begin{minipage}[t]{0.49\textwidth}
 \makebox[\textwidth][t]{
   \includegraphics[
      width=\textwidth,
      ]{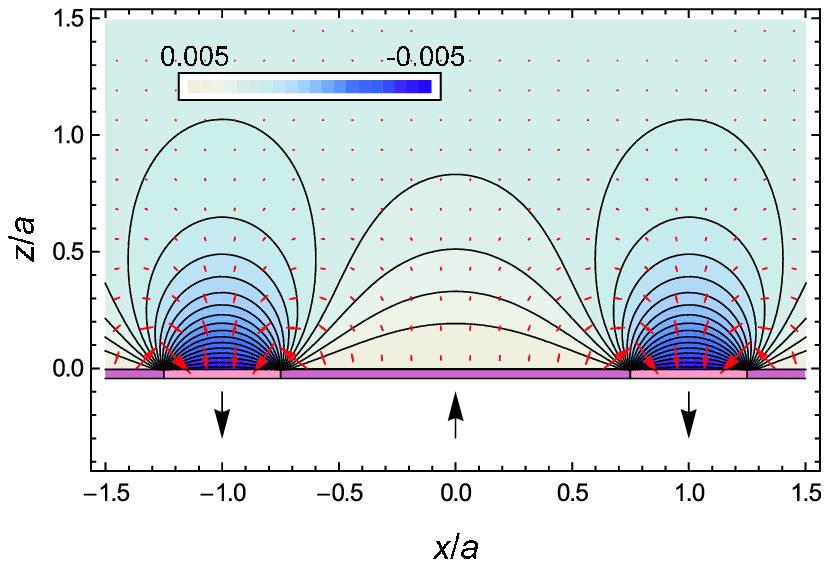}
 }
 \makebox[\textwidth][b]{\hfill {\small a) $R=0.1$ and $k\lambda_D=10^3$} \hfill}
\end{minipage}
\hfill
\begin{minipage}[t]{0.49\textwidth}
 \makebox[\textwidth][t]{
   \includegraphics[
      width=\textwidth
      ]{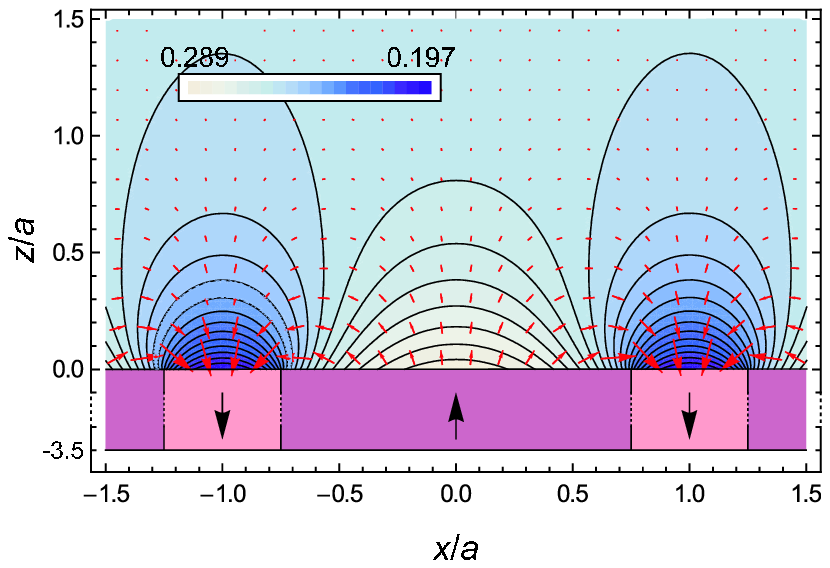}
 }
 \makebox[\textwidth][b]{\hfill {\small  b) $R=10$ and $k\lambda_D=10^3$} \hfill}
\end{minipage}
\vspace{2mm}
\end{minipage}
\begin{minipage}[t]{\textwidth}
\begin{minipage}[t]{0.49\textwidth}
 \makebox[\textwidth][t]{
   \includegraphics[
      width=\textwidth,
      ]{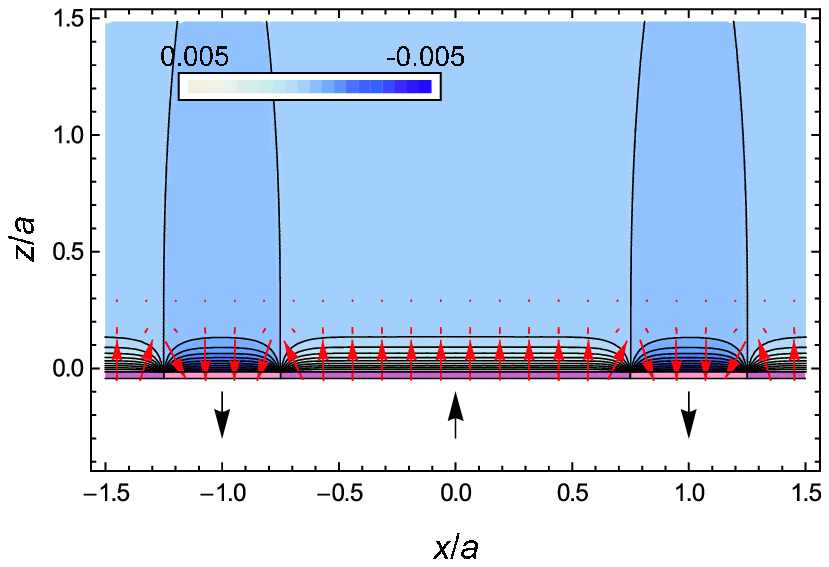}
 }
 \makebox[\textwidth][b]{\hfill {\small c) $R=0.1$ and $k\lambda_D=0.2$} \hfill}
\end{minipage}
\hfill
\begin{minipage}[t]{0.49\textwidth}
 \makebox[\textwidth][t]{
   \includegraphics[
      width=\textwidth
      ]{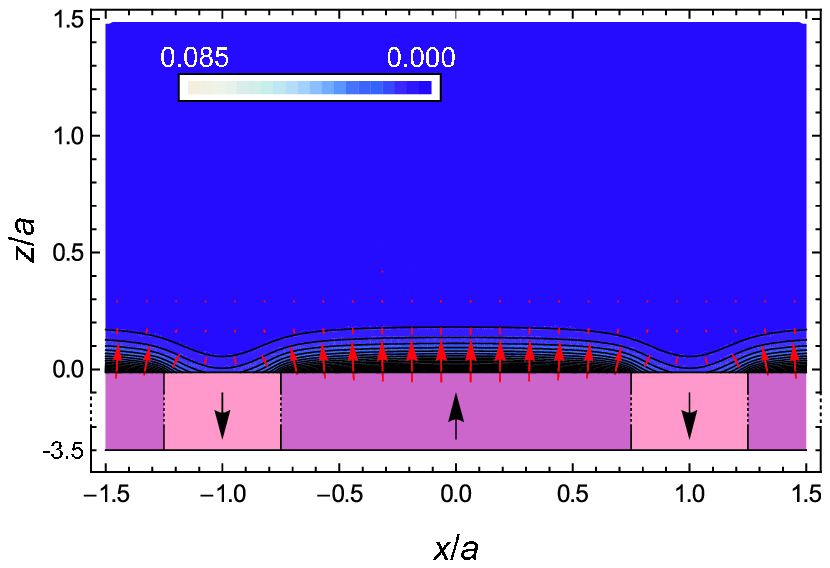}
 }
 \makebox[\textwidth][b]{\hfill {\small  d) $R=10$ and $k\lambda_D=0.2$} \hfill}
\end{minipage}
\end{minipage}
\end{minipage}
\hfill
 \begin{minipage}{0.29\textwidth}
\vfill
\caption{
Contour plot of the electrostatic potential produced by the domain pattern in 
the system with a thin ferroelectric film and wide domains, $R=0.1$, (a, c) 
and with a dense domain pattern, $R=10$, (b, d). In addition, situations with 
the absence of free charge carriers, $k\lambda_D=10^3$, (a, b), and with free 
charge screening, $k\lambda_D=0.2$, (c, d), are compared. The electrostatic 
potential is normalized to the constant ${(C_E/k)}$ given by 
Eq.~(\ref{eq04c}). Red arrows represent the inhomogeneous electric field 
produced by the domain pattern. The following values of material parameters 
were considered $\hat{p}=0.5$, $g=15.8$. Note the difference in the peak 
values of the normalized electrostatic potential and in the normalized 
electrostatic potential gradient in the vicinity of the domain walls in the 
cases (a, c) and (b, d).
\label{fig02}}
\vfill
\end{minipage}
\end{minipage}
}{
%
%
%
%
\begin{minipage}[t]{\textwidth}
\begin{minipage}[t]{0.49\textwidth}
 \makebox[\textwidth][t]{
   \includegraphics[
      width=\textwidth,
      ]{fig02a}
 }
 \makebox[\textwidth][b]{\hfill {\small a) $R=0.1$ and $k\lambda_D=10^3$} \hfill}
\end{minipage}
\hfill
\begin{minipage}[t]{0.49\textwidth}
 \makebox[\textwidth][t]{
   \includegraphics[
      width=\textwidth
      ]{fig02b}
 }
 \makebox[\textwidth][b]{\hfill {\small  b) $R=10$ and $k\lambda_D=10^3$} \hfill}
\end{minipage}
\vspace{2mm}
\end{minipage}
\begin{minipage}[t]{\textwidth}
\begin{minipage}[t]{0.49\textwidth}
 \makebox[\textwidth][t]{
   \includegraphics[
      width=\textwidth,
      ]{fig02c}
 }
 \makebox[\textwidth][b]{\hfill {\small c) $R=0.1$ and $k\lambda_D=0.2$} \hfill}
\end{minipage}
\hfill
\begin{minipage}[t]{0.49\textwidth}
 \makebox[\textwidth][t]{
   \includegraphics[
      width=\textwidth
      ]{fig02d}
 }
 \makebox[\textwidth][b]{\hfill {\small  d) $R=10$ and $k\lambda_D=0.2$} \hfill}
\end{minipage}
\end{minipage}
 \begin{minipage}{\textwidth}
\caption{
Contour plot of the electrostatic potential produced by the domain pattern in 
the system with a thin ferroelectric film and wide domains, $R=0.1$, (a, c) 
and with a dense domain pattern, $R=10$, (b, d). In addition, situations with 
the absence of free charge carriers, $k\lambda_D=10^3$, (a, b), and with free 
charge screening, $k\lambda_D=0.2$, (c, d), are compared. The electrostatic 
potential is normalized to the constant ${(C_E/k)}$ given by 
Eq.~(\ref{eq04c}). Red arrows represent the inhomogeneous electric field 
produced by the domain pattern. The following values of material parameters 
were considered $\hat{p}=0.5$, $g=15.8$. Note the difference in the peak 
values of the normalized electrostatic potential and in the normalized 
electrostatic potential gradient in the vicinity of the domain walls in the 
cases (a, c) and (b, d).
\label{fig02}}
\end{minipage}
}
\end{center}
\end{figure*}
%
%

\subsection{Systems with thin ferroelectric films and wide domains}
\label{subsec_wide_domains}

The first situation is considered for the systems with an isolating dielectric 
liquid where the thickness of the ferroelectric film $d$ is smaller than the 
average distance between the domain walls $a$, i.e. $R\ll 1$. In addition, it 
is reasonable to consider that the permittivity of ferroelectric film is much 
larger than the permittivity of the dielectric liquid, i.e. $g\gg 1$. In this 
case, the factors $(1+g\coth nR)$ in the denominator of the leading terms in 
Eq.~(\ref{eq04a}) are dominated by $g\coth nR$ and it is quite acceptable to 
approximate $(1+g\coth nR)$ by $g(1+\coth nR)$. Using the Euler formula 
$e^{iz}=\cos z + i \sin z$, using the sum and difference formulas for the 
trigonometric functions, and using the following formula:
\begin{equation}
	\sum\limits_{n=1}^\infty 
	\frac{
		e^{n(\xi + i\eta)}
	}{
		n^2\left(1 + \coth nR\right)
	}
	=
	\frac 12
	\left[
		\Li{2}\left(e^{\xi + i\eta}\right)
		-
		\Li{2}\left(e^{\xi - 2R + i\eta}\right)
	\right],
	\label{eq05}
\end{equation}
where $\Li{s}z = \sum_{n=1}^\infty z^n/n^s$ is the polylogarithm function, the 
electrostatic potential in the dielectric medium can be expressed in a form:
\begin{subequations}
\label{eq06}
\begin{equation}	
	\varphi^{(w)} = 
	\frac{C_E}{k}\,
	\Phi^{(w)}\left(
		\frac{\pi x}{a}, 
		\frac{\pi z}{a}, 
		\frac{\pi d}{2a}\sqrt{\frac{\varepsilon_a}{\varepsilon_c}}, 
		\frac{\pi P_N}{2P_0}
	\right),
	\label{eq06a}
\end{equation}
where the function $\Phi^{(w)}$ stands for the sum in Eq.~(\ref{eq04a})
\ifthenelse{\equal{\twocolumnmode}{true}}{
%
%
%
%
\begin{eqnarray}
	\label{eq06b}
	\Phi^{(w)}(\hat{x}, \hat{z}, R, \hat{p})
	&=&
	\frac{\hat{p}R}{2g}
	+
	\frac{1}{8g}
	\left[
		f(\hat{x},\hat{z},\hat{p}) +
	\right.
	\hspace{1.0cm}	
	\\ \nonumber   
	\lefteqn{
	\hspace{-2cm}
	\left. 
		f(-\hat{x},\hat{z},\hat{p}) -
		f(\hat{x},\hat{z}+2R,\hat{p}) -
		f(-\hat{x},\hat{z}+2R,\hat{p})	
	\right] 
	}
\end{eqnarray}
}{
%
%
%
%
\begin{equation}
	\label{eq06b}
	\Phi^{(w)}(\hat{x}, \hat{z}, R, \hat{p})
	=
	\frac{\hat{p}R}{2g}
	+
	\frac{1}{8g}
	\left[
		f(\hat{x},\hat{z},\hat{p}) +
		f(-\hat{x},\hat{z},\hat{p}) -
		f(\hat{x},\hat{z}+2R,\hat{p}) -
		f(-\hat{x},\hat{z}+2R,\hat{p})	
	\right] 
\end{equation}
}
and
\begin{equation}
	\label{eq06c}
	f(\xi,\eta,\zeta)
	=
	i \,\left\{
    	\Li{2}
		\left[
			-i \,e^{-\eta - i \,\left(\xi+\zeta \right) }
		\right] 
		-\Li{2}
		\left[
			i \,e^{-\eta - i \,\left(\xi-\zeta\right) }
		\right]
	\right\}.
\end{equation}
\end{subequations}
The superscript ${}^{(w)}$ indicates that the functions are 
expressed in the approximation of the thin ferroelectric film with wide 
domains. It should be noted that in the case of systems where the permittivity 
of the dielectric liquid, $\varepsilon_m$, is approximately equal to 
$\sqrt{\varepsilon_c\varepsilon_a}$, i.e. $g\approx 1$, the above expressions 
Eqs.~(\ref{eq06}) equal exactly those in Eq.~(\ref{eq04a}).

The electric field in the liquid produced by the ferroelectric 
domain pattern with wide domains can be expressed in a straightforward way:
\ifthenelse{\equal{\twocolumnmode}{true}}{
%
%
%
%
\begin{subequations}
\begin{eqnarray}
	E_x^{(w)} &=& - \frac{\partial\varphi^{(w)}}{\partial x} =
	\\ \nonumber 
	\lefteqn{
	-
	{C_E}\, 
	\Phi^{(w)}_{\hat{x}}\left(
		\frac{\pi x}{a}, 
		\frac{\pi z}{a}, 
		\frac{\pi d}{2a}\sqrt{\frac{\varepsilon_a}{\varepsilon_c}}, 
		\frac{\pi P_N}{2P_0}
	\right),
	\label{eq07a} 
	}
\end{eqnarray}
\begin{eqnarray}
	E_z^{(w)} &=& - \frac{\partial\varphi^{(w)}}{\partial z} = 
	\\ \nonumber 
	\lefteqn{
	-
	{C_E}\, 
	\Phi^{(w)}_{\hat{z}}\left(
		\frac{\pi x}{a}, 
		\frac{\pi z}{a}, 
		\frac{\pi d}{2a}\sqrt{\frac{\varepsilon_a}{\varepsilon_c}}, 
		\frac{\pi P_N}{2P_0}
	\right),
	\label{eq07b}
	} 
\end{eqnarray}
\end{subequations}
}{
%
%
%
%
%
\begin{subequations}
\begin{eqnarray}
	E_x^{(w)} &=& - \frac{\partial\varphi^{(w)}}{\partial x} = 
	-
	{C_E}\, 
	\Phi^{(w)}_{\hat{x}}\left(
		\frac{\pi x}{a}, 
		\frac{\pi z}{a}, 
		\frac{\pi d}{2a}\sqrt{\frac{\varepsilon_a}{\varepsilon_c}}, 
		\frac{\pi P_N}{2P_0}
	\right),
	\label{eq07a} \\ 
	E_z^{(w)} &=& - \frac{\partial\varphi^{(w)}}{\partial z} = 
	-
	{C_E}\, 
	\Phi^{(w)}_{\hat{z}}\left(
		\frac{\pi x}{a}, 
		\frac{\pi z}{a}, 
		\frac{\pi d}{2a}\sqrt{\frac{\varepsilon_a}{\varepsilon_c}}, 
		\frac{\pi P_N}{2P_0}
	\right),
	\label{eq07b} 
\end{eqnarray}
\end{subequations}
}
\ifthenelse{\equal{\twocolumnmode}{true}}{
%
%
%
%
%
\begin{widetext}
where 
\begin{subequations}
\label{eq08}
\begin{eqnarray}
	\Phi^{(w)}_{\hat{x}}\left(\hat{x}, \hat{z}, R, \hat{p}\right)
	&=&
	\frac {1}{8g}
	\log 
	\frac{
		\left[
			\cosh\hat{z} - \sin(\hat{x}-\hat{p}) 
		\right] 
     	\left[
			\cosh(\hat{z}+2R) + \sin(\hat{x}+\hat{p}) 
		\right]
	}{
     	\left[ 
			\cosh\hat{z} + \sin(\hat{x}+\hat{p}) 
		\right] 
     	\left[ 
			\cosh(\hat{z}+2R) - \sin(\hat{x}-\hat{p}) 
		\right]
	},		 
	\label{eq08a}
\end{eqnarray}
\begin{eqnarray}
	\Phi^{(w)}_{\hat{z}}
	\left(\hat{x}, \hat{z}, R, \hat{p}\right)
	&=& 
	-\frac{1}{4g}
	\arctan 
	\frac{
		2\cos\hat{p}\,
		\left( 
			e^{\hat{z}}\,\cos\hat{x} + \sin\hat{p} 
		\right) 
	}{
		e^{\hat{z}}\,
		\left( 
			e^{\hat{z}} + 2\,\cos\hat{x}\,\sin\hat{p} 
		\right)
		-\cos 2\hat{p} 
	} + 
	\label{eq08b}
	\\
	\lefteqn{
	\frac{1}{4g}
	\arctan 
	\frac{
		2\cos\hat{p}
		\left( 
			\cos\hat{x} + e^{-\hat{z}-2R}\,\sin\hat{p}
		\right) 
	}{
		2\sin\hat{p}\,
		\left[ 
			\cos\hat{x} + \cosh (\hat{z}+2R)\,\sin\hat{p} 
		\right]  + 
		2\cos^2\hat{p}\,\sinh (\hat{z}+2R)
	}.
	}
	\nonumber
\end{eqnarray}
\end{subequations}
\end{widetext}
}{
%
%
%
%
%
where
\begin{subequations}
\begin{eqnarray}
	\Phi^{(w)}_{\hat{x}}\left(\hat{x}, \hat{z}, R, \hat{p}\right)
	&=&
	\frac {1}{8g}
	\log 
	\frac{
		\left[
			\cosh\hat{z} - \sin(\hat{x}-\hat{p}) 
		\right] 
     	\left[
			\cosh(\hat{z}+2R) + \sin(\hat{x}+\hat{p}) 
		\right]
	}{
     	\left[ 
			\cosh\hat{z} + \sin(\hat{x}+\hat{p}) 
		\right] 
     	\left[ 
			\cosh(\hat{z}+2R) - \sin(\hat{x}-\hat{p}) 
		\right]
	},		 
	\label{eq08a}\\
	\Phi^{(w)}_{\hat{z}}
	\left(\hat{x}, \hat{z}, R, \hat{p}\right)
	&=& 
	-\frac{1}{4g}
	\arctan 
	\frac{
		2\cos\hat{p}\,
		\left( 
			e^{\hat{z}}\,\cos\hat{x} + \sin\hat{p} 
		\right) 
	}{
		e^{\hat{z}}\,
		\left( 
			e^{\hat{z}} + 2\,\cos\hat{x}\,\sin\hat{p} 
		\right)
		-\cos 2\hat{p} 
	} + 
	\nonumber \\
	\lefteqn{
	\hspace{-1cm} \frac{1}{4g}
	\arctan 
	\frac{
		2\cos\hat{p}
		\left( 
			\cos\hat{x} + e^{-\hat{z}-2R}\,\sin\hat{p}
		\right) 
	}{
		2\sin\hat{p}\,
		\left[ 
			\cos\hat{x} + \cosh (\hat{z}+2R)\,\sin\hat{p} 
		\right]  + 
		2\cos^2\hat{p}\,\sinh (\hat{z}+2R)
	}.
	} 	\label{eq08b}
\end{eqnarray}
\end{subequations}
}
In the derivation of Eqs.~(\ref{eq08a}), we have used the property 
of the polylogarithm function $\partial(\Li{2} z)/\partial z = -\log(1-z)/z$ 
and the definition of the logarithm in a complex plane $\log z=\log|z| + i\arg 
z$.

\subsection{Systems with dense domain patterns}
\label{subsec_dense_domains}

The second situation, for which it is possible to write down a closed formula 
for the electrostatic potential and the components of the electric field, is 
the ferroelectric film with a dense domain pattern, i.e. the systems where the 
thickness of the ferroelectric film $d$ is larger than the average distance 
between the domain walls $a$, i.e. $R\gg 1$. In this case, the terms $\coth 
nR$ in the denominator of the leading terms in Eq.~(\ref{eq04a}) can be 
replaced by 1 and, thus, it is quite acceptable to approximate $(1+g\coth nR)$ 
by $(1+g)$. Following the same algebraic procedures as in the previous 
subsection, the electrostatic potential in the dielectric liquid can be 
expressed in a form:
\begin{subequations}
\begin{equation}	
	\varphi^{(d)} = 
	\frac{C_E}{k}\, 
	\Phi^{(d)}\left(
		\frac{\pi x}{a}, 
		\frac{\pi z}{a}, 
		\frac{\pi P_N}{2P_0}
	\right),
	\label{eq09a}
\end{equation}
where the function $\Phi^{(d)}$ stands for the sum in Eq.~(\ref{eq04a})
\ifthenelse{\equal{\twocolumnmode}{true}}{
%
%
%
%
\begin{equation}
	\label{eq09b}
	\Phi^{(d)}(\hat{x}, \hat{z}, \hat{p})
	=
	\frac{\hat{p}R}{2g}
	+ 
	\frac{1}{4\left(1+g\right)}
	\left[
		f(\hat{x},\hat{z},\hat{p}) 
		+ f(-\hat{x},\hat{z},\hat{p}) 
	\right].
\end{equation}
}{
%
%
%
%
%
\begin{equation}
	\label{eq09b}
	\Phi^{(d)}(\hat{x}, \hat{z}, \hat{p})
	=
	\frac{\hat{p}R}{2g}
	+ 
	\frac{1}{4\left(1+g\right)}
	\left[
		f(\hat{x},\hat{z},\hat{p}) 
		+ f(-\hat{x},\hat{z},\hat{p}) 
	\right].
\end{equation}
}
\end{subequations}
The superscript ${}^{(d)}$ indicates that the functions are 
expressed in the approximation of the dense domain pattern. 
 
The electric field in the dielectric liquid produced by the dense 
ferroelectric domain pattern can be expressed in a straightforward way:
\ifthenelse{\equal{\twocolumnmode}{true}}{
%
%
%
\begin{subequations}
\begin{eqnarray}
	\hspace{-8mm}
	E_x^{(d)} &=& - \frac{\partial\varphi^{(d)}}{\partial x} = 
	-
	{C_E}\,
	\Phi^{(d)}_{\hat{x}}\left(
		\frac{\pi x}{a}, 
		\frac{\pi z}{a}, 
		\frac{\pi P_N}{2P_0}
	\right),
	\label{eq10a} \\ 
	\hspace{-8mm}
	E_z^{(d)} &=& - \frac{\partial\varphi^{(d)}}{\partial z} = 
	-
	{C_E}\,
	\Phi^{(d)}_{\hat{z}}\left(
		\frac{\pi x}{a}, 
		\frac{\pi z}{a}, 
		\frac{\pi P_N}{2P_0}
	\right),
	\label{eq10b} 
\end{eqnarray}
\end{subequations}
}{
%
%
%
%
\begin{subequations}
\begin{eqnarray}
	E_x^{(d)} &=& - \frac{\partial\varphi^{(d)}}{\partial x} = 
	-
	{C_E}\,
	\Phi^{(d)}_{\hat{x}}\left(
		\frac{\pi x}{a}, 
		\frac{\pi z}{a}, 
		\frac{\pi P_N}{2P_0}
	\right),
	\label{eq10a} \\ 
	E_z^{(d)} &=& - \frac{\partial\varphi^{(d)}}{\partial z} = 
	-
	{C_E}\,
	\Phi^{(d)}_{\hat{z}}\left(
		\frac{\pi x}{a}, 
		\frac{\pi z}{a}, 
		\frac{\pi P_N}{2P_0}
	\right),
	\label{eq10b} 
\end{eqnarray}
\end{subequations}
}
where
\ifthenelse{\equal{\twocolumnmode}{true}}{
%
%
%
\begin{subequations}
\begin{eqnarray}
	\Phi^{(d)}_{\hat{x}}\left(\hat{x}, \hat{z}, \hat{p}\right)
	&=&
	\frac {1}{4\left(1+g\right)}
	\log 
	\frac{
		\cosh\hat{z} - \sin(\hat{x}-\hat{p}) 
	}{
		\cosh\hat{z} + \sin(\hat{x}+\hat{p}) 
	}.\hspace{1cm}		 
	\label{eq11a}\\
	\Phi^{(d)}_{\hat{z}}
	\left(\hat{x}, \hat{z}, \hat{p}\right)
	&=& 
	-\frac {1}{2\left(1+g\right)}
	\times
	\label{eq11b}
	\\ \nonumber
	\lefteqn{
	\arctan 
	\frac{
		2\cos\hat{p}\,
		\left( 
			e^{\hat{z}}\,\cos\hat{x} + \sin\hat{p} 
		\right) 
	}{
		e^{\hat{z}}\,
		\left( 
			e^{\hat{z}} + 2\,\cos\hat{x}\,\sin\hat{p} 
		\right)
		-\cos 2\hat{p} 
	}. 
	}  
\end{eqnarray}
\end{subequations}
}{
%
%
%
%
\begin{subequations}
\begin{eqnarray}
	\Phi^{(d)}_{\hat{x}}\left(\hat{x}, \hat{z}, \hat{p}\right)
	&=&
	\frac {1}{4\left(1+g\right)}
	\log 
	\frac{
		\cosh\hat{z} - \sin(\hat{x}-\hat{p}) 
	}{
		\cosh\hat{z} + \sin(\hat{x}+\hat{p}) 
	}.		 
	\label{eq11a}\\
	\Phi^{(d)}_{\hat{z}}
	\left(\hat{x}, \hat{z}, \hat{p}\right)
	&=& 
	-\frac {1}{2\left(1+g\right)}
	\arctan 
	\frac{
		2\cos\hat{p}\,
		\left( 
			e^{\hat{z}}\,\cos\hat{x} + \sin\hat{p} 
		\right) 
	}{
		e^{\hat{z}}\,
		\left( 
			e^{\hat{z}} + 2\,\cos\hat{x}\,\sin\hat{p} 
		\right)
		-\cos 2\hat{p} 
	}.  
	\label{eq11b}
\end{eqnarray}
\end{subequations}
}

\subsection{Simulation of the electrostatic potential and electrophoretic force}
\label{subsec_simulation}

Figures \ref{fig02}a and \ref{fig02}b show the spatial distribution of the 
electrostatic potential $\varphi$ given by Eq.~(\ref{eq04a}) and the vectors 
of the electric field produced by the periodic domain pattern in the 
dielectric liquid in the two aforementioned situations characterized by the 
absence of the free charge carriers in the liquid medium (i.e. with a large 
value of the Debye screening length $k\lambda_D=10^3$). Figure \ref{fig02}a 
shows the system with a thin ferroelectric film and wide domains, $R=0.1$, and 
Figure \ref{fig02}b shows the system with a dense domain pattern, $R=10$, (b). 
The electrostatic potential $\varphi$ is normalized to the constant $({C_E/k})
$ given by Eq.~(\ref{eq04c}) and, thus, the contour plots correspond to the 
spatial distributions of the dimensionless functions $\Phi$ given by 
Eq.~(\ref{eq06b}) and (\ref{eq09b}). Figures \ref{fig02}c and \ref{fig02}d 
present the effect of the inhomogeneous electric field screening by free 
charge carriers in the ionic liquid (i.e. the small value of the Debye 
screening length compared to the domain spacing, $k\lambda_D=0.2$) in the 
systems with the same geometry as in Figs. \ref{fig02}a and \ref{fig02}b. In 
the bottom part of the figure, there are indicated the positions of the 
180${}^\circ$ domain walls and the thickness of the ferroelectric film. Red 
arrows represents the vectors of the inhomogeneous electric field produced by 
the domain pattern and they are proportional to the electrophoretic (EP) force,
 i.e. the Coulomb force acting on charged particles.

If we consider an approximately spherical particle of a diameter $r$ with free 
charges on its surface of a surface density $\sigma$, the electrophoretic 
force acting on such a particle is given by the classical formula:
\begin{equation}
	F_{ep} = - 4\pi r^2 \sigma \nabla \varphi.
	\label{eq12} 
\end{equation}
With use of the notation adopted in this Article, Eq.~(\ref{eq12}) can be 
rewritten in the form:
\begin{subequations}
\begin{eqnarray}
	F_{ep} &=& 
		- 4\pi r^2 \sigma \left(C_E/k\right) \nabla \Phi
		\label{eq13a} \\
	&=& 
		- C_{ep} \widehat{\nabla} \Phi,
	\nonumber
\end{eqnarray}  
where
\begin{equation}
	C_{ep} = 4\pi r^2 \sigma C_E
	\label{eq13b}
\end{equation}
\end{subequations}
and the symbol $\widehat{\nabla}$ stands for the gradient operator in the 
$\hat{x}_i$-coordinate system, where $\hat{x}_i=kx_i$.

%
%
\begin{figure*}[t]
\begin{center}
\ifthenelse{\equal{\twocolumnmode}{true}}{
%
%
%
\begin{minipage}[t]{\textwidth}
\begin{minipage}[t]{0.69\textwidth}
\begin{minipage}[t]{\textwidth}
\begin{minipage}[t]{0.49\textwidth}
 \makebox[\textwidth][t]{
   \includegraphics[
      width=\textwidth,
      ]{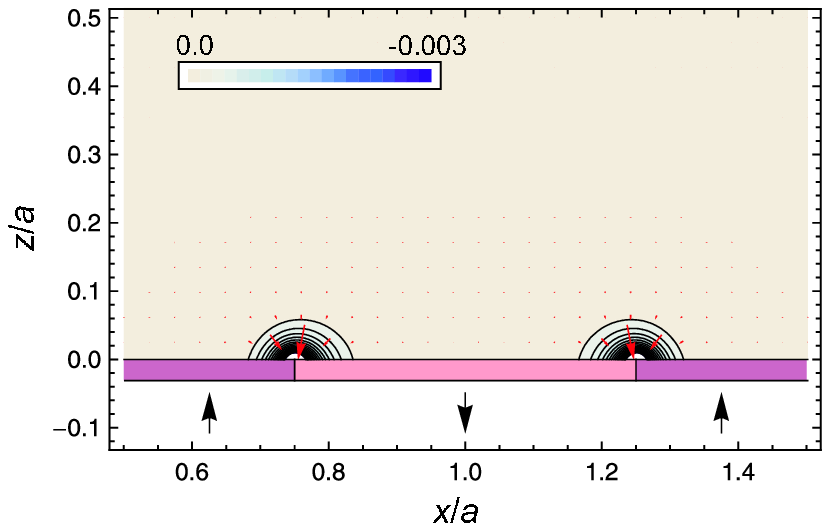}
 }
 \makebox[\textwidth][b]{\hfill {\small a) $R=0.1$ and $k\lambda_D=10^3$} \hfill}
\end{minipage}
\hfill
\begin{minipage}[t]{0.49\textwidth}
 \makebox[\textwidth][t]{
   \includegraphics[
      width=\textwidth
      ]{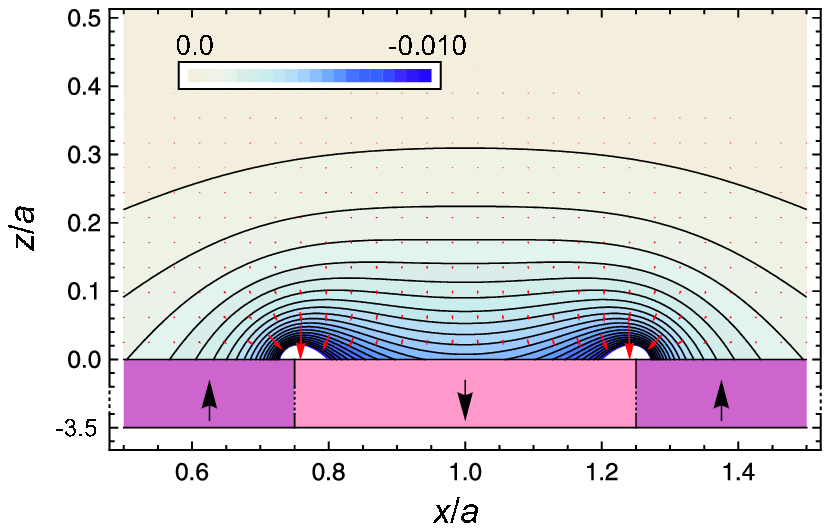}
 }
 \makebox[\textwidth][b]{\hfill {\small  b) $R=10$ and $k\lambda_D=10^3$} \hfill}  
\end{minipage}
\vspace{2mm}
\end{minipage}
\begin{minipage}[t]{\textwidth}
\begin{minipage}[t]{0.49\textwidth}
 \makebox[\textwidth][t]{
   \includegraphics[
      width=\textwidth,
      ]{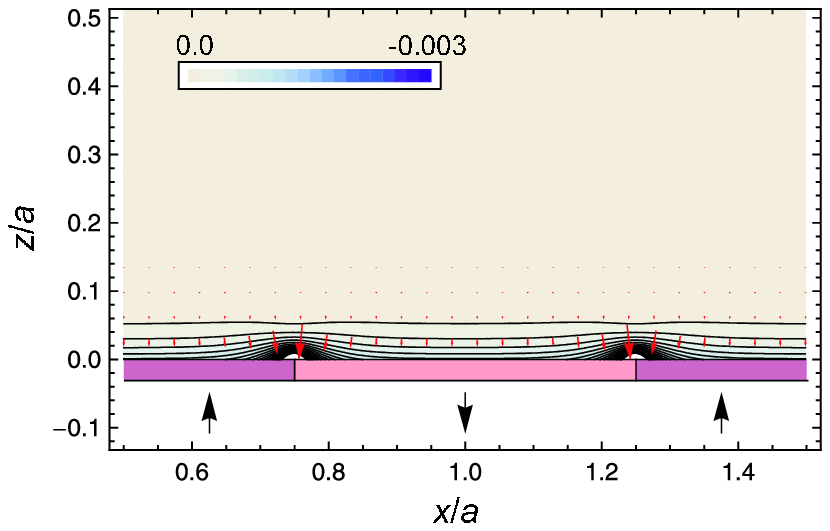}
 }
 \makebox[\textwidth][b]{\hfill {\small c) $R=0.1$ and $k\lambda_D=0.2$} \hfill}
\end{minipage}
\hfill
\begin{minipage}[t]{0.49\textwidth}
 \makebox[\textwidth][t]{
   \includegraphics[
      width=\textwidth
      ]{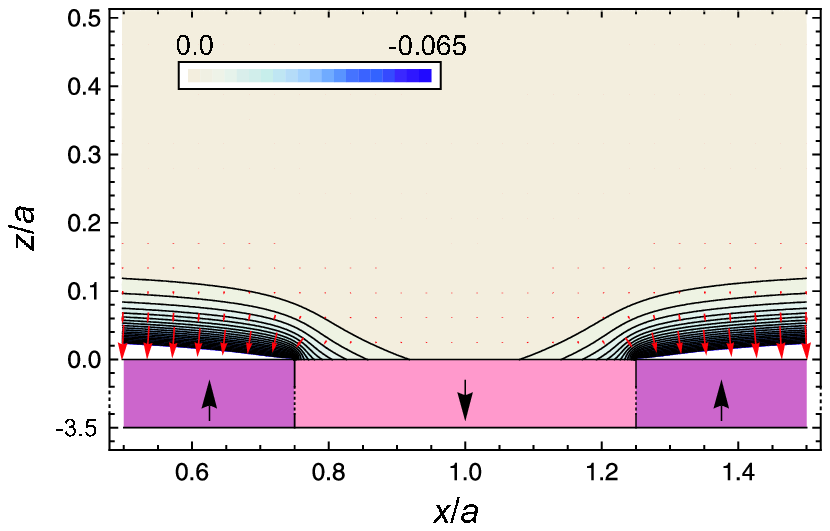}
 }
 \makebox[\textwidth][b]{\hfill {\small  d) $R=10$ and $k\lambda_D=0.2$} \hfill} 
\end{minipage}
\end{minipage}
\end{minipage}
\hfill
 \begin{minipage}{0.29\textwidth}
\vfill
\caption{
Contour plot of the positive DEP force potential produced by the domain 
pattern in the system with a thin ferroelectric film and wide domains, $R=0.1$,
 (a, c) and with a dense domain pattern, $R=10$, (b, d). In addition, 
situations with the absence of free charge carriers, $k\lambda_D=10^3$, (a, b)
, and with free charge screening, $k\lambda_D=0.2$, (c, d), are compared. The 
potential is normalized to $C_{dep}$. Red arrows represents the DEP force 
produced by the domain pattern. The following values of material parameters 
were considered $\hat{p}=0.5$, $g=15.8$. Note the difference in the peak 
values of the normalized dielectrophoretic force potential in the vicinity of 
the domain walls in the cases (a, c) and (b, d), which is controlled only by 
the geometry of the domain pattern.
\label{fig03}}
\vfill
\end{minipage}
\end{minipage}
}{
%
%
%
%
\begin{minipage}[t]{\textwidth}
\begin{minipage}[t]{0.49\textwidth}
 \makebox[\textwidth][t]{
   \includegraphics[
      width=\textwidth,
      ]{fig03a}
 }
 \makebox[\textwidth][b]{\hfill {\small a) $R=0.1$ and $k\lambda_D=10^3$} \hfill}
\end{minipage}
\hfill
\begin{minipage}[t]{0.49\textwidth}
 \makebox[\textwidth][t]{
   \includegraphics[
      width=\textwidth
      ]{fig03b}
 }
 \makebox[\textwidth][b]{\hfill {\small  b) $R=10$ and $k\lambda_D=10^3$} \hfill}  
\end{minipage}
\vspace{2mm}
\end{minipage}
\begin{minipage}[t]{\textwidth}
\begin{minipage}[t]{0.49\textwidth}
 \makebox[\textwidth][t]{
   \includegraphics[
      width=\textwidth,
      ]{fig03c}
 }
 \makebox[\textwidth][b]{\hfill {\small c) $R=0.1$ and $k\lambda_D=0.2$} \hfill}
\end{minipage}
\hfill
\begin{minipage}[t]{0.49\textwidth}
 \makebox[\textwidth][t]{
   \includegraphics[
      width=\textwidth
      ]{fig03d}
 }
 \makebox[\textwidth][b]{\hfill {\small  d) $R=10$ and $k\lambda_D=0.2$} \hfill} 
\end{minipage}
\end{minipage}
\begin{minipage}{\textwidth}
\caption{
Contour plot of the positive DEP force potential produced by the domain 
pattern in the system with a thin ferroelectric film and wide domains, $R=0.1$,
 (a, c) and with a dense domain pattern, $R=10$, (b, d). In addition, 
situations with the absence of free charge carriers, $k\lambda_D=10^3$, (a, b)
, and with free charge screening, $k\lambda_D=0.2$, (c, d), are compared. The 
potential is normalized to $C_{dep}$. Red arrows represents the DEP force 
produced by the domain pattern. The following values of material parameters 
were considered $\hat{p}=0.5$, $g=15.8$. Note the difference in the peak 
values of the normalized dielectrophoretic force potential in the vicinity of 
the domain walls in the cases (a, c) and (b, d), which is controlled only by 
the geometry of the domain pattern.
\label{fig03}}
\end{minipage}
}
\end{center}
\end{figure*}
%
%
%
In the numerical calculation of the normalized electrostatic potential, we 
considered the values of the ionic liquid permittivity $\varepsilon_m=3$ and 
the in-plane and out-of plane permittivity of the ferroelectric 
$\varepsilon_a=84$ and $\varepsilon_a=29$, respectively, which corresponds to 
the dielectric properties of lithium niobate 
\cite{Savage.JApplPhys.37.1966,Scrymgeour.PhysRevB.71.2005}. The three 
aforementioned values of permittivity give the value of the parameter 
$g=15.8$. Finally, the non-neutral domain pattern was considered with 
$\hat{p}=0.5$, which corresponds to $P_N/P_0=0.32$. It is noticeable to see 
the difference in the peak values of the normalized electrostatic potential 
-0.005 and 0.005 in Fig.~\ref{fig02}a; and 0.197 and 0.289 in 
Fig.~\ref{fig02}b, which is not controlled by the values of spontaneous 
polarization $P_0$ but by the ratio of the ferroelectric film thickness over 
the domain spacing $d/a$, which is proportional to the value of the parameter 
$R$. The similar effect of the value of $R$ can bee seen on the normalized 
electrostatic potential gradient in the vicinity of the domain walls in 
Figs.~\ref{fig02}a and \ref{fig02}b.


\section{Dielectrophoretic force}
\label{sec_DEP}

The dielectrophoretic force $F_{dep}$ that acts on approximately spherical 
particle of a diameter $r$ and the dielectric constant $\varepsilon_p$ in the 
inhomogeneous electric field $E$ is given by the classical 
formula\cite{Pohl.JApplPhys.22.1951}:
\begin{subequations}
\begin{equation}
	F_{dep} = K_{dep} \nabla\left(E_x^2 + E_z^2\right)
	\label{eq14a},
\end{equation}  
where
\begin{equation}
	K_{dep} = 2\pi r^3 \varepsilon_m\varepsilon_0\, 
	\frac{
		\varepsilon_p - \varepsilon_m
	}{	
		\varepsilon_p + 2\varepsilon_m
	}
	\label{eq14b}.
\end{equation}  
\end{subequations}
It should be noted that depending on the permittivity values of the medium and 
particles, the situation characterized by $\varepsilon_p > \varepsilon_m$, 
i.e. $K_{dep}>0$, is called positive dielectrophoresis and the situation, when 
$\varepsilon_p < \varepsilon_m$, i.e. $K_{dep}<0$, is called negative 
dielectrophoresis.

With use of the notation adopted in this Article, Eq.~(\ref{eq14a}) can be 
rewritten in the form:
\begin{eqnarray}
	F_{dep} &=& 
		K_{dep}\,C_E^2\,
		\nabla
		\left(\Phi_{\hat{x}}^2 + \Phi_{\hat{z}}^2\right)
	\label{eq15} \\
	&=& 
		K_{dep}\,C_E^2 k\,
		\widehat{\nabla}
		\left(\Phi_{\hat{x}}^2 + \Phi_{\hat{z}}^2\right).
	\nonumber
\end{eqnarray}  
For the purpose of the graphical presentation of spatial distribution of the 
DEP forces generated by the domain pattern, it is convenient to introduce the 
DEP force potential by formula $F_{dep}=-\widehat{\nabla} U_{dep}$. Thus, the 
function $U_{dep}$ can be expressed in the form
\begin{subequations}
\begin{equation}
	U_{dep} = - C_{dep}\, \left(\Phi_{\hat{x}}^2 + \Phi_{\hat{z}}^2\right)
	\label{eq16a},
\end{equation}  
where
\begin{equation}
	C_{dep} = 
		\frac{
			32\,P_0^2\,r^3\,\left(\varepsilon_p-\varepsilon_m\right) 
		}{
			a \,\varepsilon_0\,\varepsilon_m\,
			\left(\varepsilon_p+2\varepsilon_m \right) 
		}
	\label{eq16b}.
\end{equation}  
\end{subequations}

Figures \ref{fig03}a-d show the spatial distribution of the dielectrophoretic 
field potential $U_{dep}$ given by Eq.~(\ref{eq16a}) and the vectors of the 
dielectrophoretic force produced by the periodic domain pattern in the ionic 
liquid in the same four situations as in Figs.~\ref{fig02}. 
Figures~\ref{fig03}a and \ref{fig03}c show the systems with a thin 
ferroelectric film and wide domains, $R=0.1$. On the other hand, 
Figures~\ref{fig03}b and \ref{fig03}d show the systems with a dense domain 
pattern, $R=10$. The dielectrophoretic field potential $U_{dep}$ is normalized 
to the constant ${C_{dep}}$ given by Eq.~(\ref{eq16b}) and, thus, the contour 
plots correspond to the spatial distributions of the dimensionless functions 
$-\left(\Phi_{\hat{x}}^2 + \Phi_{\hat{z}}^2\right)$. Figures~\ref{fig03}a and 
\ref{fig03}b show the systems with a dielectric liquid and a low concentration 
of free charge carriers, which corresponds to the large value of the Debye 
screening length, $k\lambda_D=10^3$. On the contrary, Figs.~\ref{fig03}c and 
\ref{fig03}d show the systems, where the inhomogeneous electric field is 
screened due to free charge carriers in the ionic liquid and where the Debye 
screening length is smaller than the domain spacing, i.e. $k\lambda_D=0.2$. It 
is noticeable to see the difference in the peak values of the normalized 
dielectrophoretic field potential in Figs.~\ref{fig03}a and \ref{fig03}c, and 
\ref{fig03}b and \ref{fig03}d, respectively. Again the point is that the peak 
values are not controlled by the values of spontaneous polarization $P_0$ but 
by the ratio of the ferroelectric film thickness over the domain spacing $d/a$,
 which is proportional to the value of the parameter $R$.

\section{Discussion and conclusions}
\label{sec_discussion}

Theoretical results presented in the above Sections indicate qualitative 
features that play the key role in the decoration of nanoparticles using 
polydomain ferroelectric films. In the following subsections, we will briefly 
discuss (i) the effect of electrical neutrality of dielectric nanoparticles on 
the mechanism of nanopatterning, (ii) the effect of free charge screening on 
the inhomogeneous field and the value of DEP force, (iii) the role of domain 
pattern geometry on the value of the DEP force produced by the ferroelectric 
domain pattern, (iv) the role of the Brownian motion on the stability of the 
nanoparticle patterns, and, finally, (v) we will estimate the values of 
important numerical parameters that are essential in the design of a 
ferroelectric system used for patterning of nanoparticles.

\subsection{The role of electroneutrality of dielectric nanoparticles}
\label{subsec_electroneutrality}

Since the electrical neutrality of nanoparticles is rarely the case in real 
experimental situations, it is necessary to compare the EP and DEP forces and 
to analyze the conditions that controls switching between the EP and DEP 
mechanisms of nanoparticle deposition.

%
%
\begin{figure}[t]
\begin{center}
	\includegraphics[width=85mm]{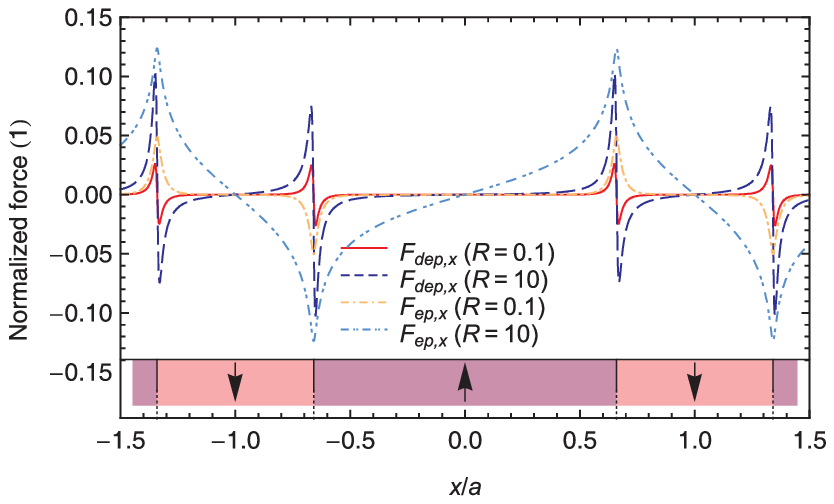}
\end{center}
\caption{
Comparison of the $x$-component of the EP and DEP force dependences on the 
ratio $x/a$ in the system with a thin ferroelectric film and wide domains, 
$R=0.1$ (solid line), and in the system with a dense domain pattern, $R=10$ 
(dashed line), at the distance $z=0.01\,a$ from the surface of the 
ferroelectric film. The EP force value is normalized to $C_{ep}$ given by 
Eq.~(\ref{eq13b}). DEP force value is normalized to $C_{dep}$ given by 
Eq.~(\ref{eq16b}) and, therefore, corresponds to the positive DEP force. Note 
the qualitative difference between the EP and DEP force spatial distribution.
\label{fig04}}
\end{figure}
%
%
%
Figure~\ref{fig04} shows the comparison of the EP and DEP force dependences on 
the ratio $x/a$ in the system with a thin ferroelectric film and wide domains, 
$R=0.1$ (solid line), and in the system with a dense domain pattern, $R=10$ 
(dashed line), at the distance $z=0.01\,a$ from the surface of the 
ferroelectric film. The EP force value is normalized to $C_{ep}$ given by 
Eq.~(\ref{eq13b}). DEP force value is normalized to $C_{dep}$ given by 
Eq.~(\ref{eq16b}) and, therefore, corresponds to the positive DEP force. It is 
seen that in the case of the EP mechanism, the $x$-component of the EP force 
is zero in the middle of each domain and the potentially charged nanoparticles 
would be preferentially deposited in the middle of those domains that have the 
opposite charge than the charged nanoparticle. On the contrary, the 
$x$-component of the DEP force is zero at the domain wall position, abruptly 
increases with the increasing distance from the domain wall and reaches its 
maximum $F_{x,max}$ at the distance $\Delta x_{max}$ from the domain wall. It 
is seen that the reduction of the thickness of the film by a factor of 100, 
the peak values of the DEP force $x$-component a smaller by a factor of 10. It 
is essential that the area of the intersection of the domain wall with the 
surface of the ferroelectric film represents a {\em trap} for nanoparticles in 
the case of a positive dielectrophoresis.

Figure \ref{fig04} shows that the peak normalized values of the EP and DEP 
forces are of the same order of magnitude. Therefore, one should compare the 
parameters $C_{ep}$ and $C_{dep}$ to get the critical value of the surface 
charge that controls the switching between the EP and DEP nanoparticle 
deposition mechanism. As a result, the DEP mechanism of deposition is dominant 
over the EP mechanism, when the surface charge density of the nanoparticles is 
much smaller than the critical value:
\begin{equation}
	\sigma \ll \sigma_{crit} = 
		\frac{
			2P_0r \left(\varepsilon_p - \varepsilon_m\right)
		}{
			a \left(2\varepsilon_m + \varepsilon_p\right)
		}.
	\label{eq17}
\end{equation}

%
%
\begin{table}
\caption{Material parameters (spontaneous polarization $P_0$, permittivity 
tensor components in the direction parallel and perpendicular to the direction 
of the spontaneous polarization $\varepsilon_c$ and $\varepsilon_a$, 
respectively, and the pyroelectric coefficient $\gamma_S$ at the room 
temperature) of selected ferroelectric materials: 
LiTaO${}_3$\cite{Whatmore.RepProgPhys.49.1986,Scrymgeour.PhysRevB.71.2005}, 
LiNbO${}_3$\cite{Savage.JApplPhys.37.1966,Scrymgeour.PhysRevB.71.2005}, 
BaTiO${}_3$\cite{Chynoweth.JApplPhys.27.1956,Li.JApplPhys.98.2005}, and 
PbTiO${}_3$\cite{Haun.JApplPhys.62.1987}.
    \label{tab01}}
\begin{ruledtabular}
\begin{tabular}{llcccc}
Parameter& Unit& LiTaO${}_3$& LiNbO${}_3$& BaTiO${}_3$& PbTiO${}_3$ \cr \hline
$P_0$& C\,m${}^{-2}$& 0.55& 0.75& 0.26& 0.76 \cr
$\varepsilon_c$& 1& 44& 29& 188& 66 \cr
$\varepsilon_a$& 1& 53& 84& 3600& 124 \cr
$\gamma_S$& $\rm\mu C$\,m${}^{-2}$\,K${}^{-1}$& 230& 40& 200& 380 
\end{tabular}
\end{ruledtabular}
\end{table}
%
%
%
Table~\ref{tab01} presents numerical values of materials parameters considered 
in our theoretical analysis for lithium tantalate (LiTaO${}_3$), lithium 
niobate (LiNbO${}_3$), barium titanate (BaTiO${}_3$), and lead titanate 
(PbTiO${}_3$) that can be prepared in a form of single crystal samples. In our 
numerical estimates, we further adopt parameters of materials that have been 
used previously in experiments by Ke {\em et al.}\cite{Ke.JApplPhys.101.2007}, 
In particular, we consider the use of lithium niobate single crystal with the 
domain spacing $a=5\,{\rm \mu m}$, polystyrene nanoparticles with dielectric 
constant $\varepsilon_p=2.7$ and radius $r=65\, {\rm nm}$, and dodecane liquid 
with $\varepsilon_m=2$, the critical value of the nanoparticle surface charge 
is $\sigma_{crit}\approx 0.002\, {\rm C\,m^{-2}}$.

\subsection{The role of free charge screening}
\label{subsec_screening}

Since the nanoparticle deposition experiments are carried out in liquid media, 
the presence of free charge carriers in the liquid cannot be completely 
avoided and sometimes it is even induced intentionally. It means that the effect 
of finite Debye screening length on the values of EP and DEP forces should be 
carefully analyzed. At first let us consider the simplest case of water. At 
room temperature, the permittivity of water is about 80 and the ionic strength 
equals $I=10^{-4}\, {\rm mol\, m^{-3}}$, which yields the typical value of the 
Debye screening length $\lambda_D\approx 1\, {\rm \mu m}$. In the case of 
nonpolar dielectric liquids, such as pentanoic acid or silicone oil, the value 
of the Debye screening length can be of several orders of magnitude larger.

Figure \ref{fig02} clearly shows that the inhomogeneous electric field 
essentially decays within the distance of about $2a$ from the surface of the 
ferroelectric film. This means that the decoration of nanoparticles using DEP 
forces should not be affected by free charge carriers in the systems where the 
Debye screening length is larger than the domain period $2a$. From this 
condition, one can calculate the critical ionic strength of the liquid medium:
\begin{equation}
	I < I_{crit} = \frac{
		\varepsilon_0 \varepsilon_m k_B T
	}{
		8N_A e^2 a^2
	}.
	\label{eq18}
\end{equation}     
Considering the experiments by Grilli\cite{Grilli.ApplPhysLett.92.2008} with 
pentanoic acid $\varepsilon_m=2.66$ and the value of domain spacing $a=100\, 
{\rm \mu m}$, the Eq.~(\ref{eq18}) yields the value $I_{crit}=7.7\,10^{-7} 
{\rm mol\, m^{-3}}$.

\subsection{The role of domain pattern geometry}
\label{subsec_geometry}

%
%
\begin{figure*}[t]
\begin{center}
\begin{minipage}[t]{\textwidth}
\begin{minipage}[t]{0.49\textwidth}
 \makebox[\textwidth][t]{
   \includegraphics[
      width=\textwidth,
      ]{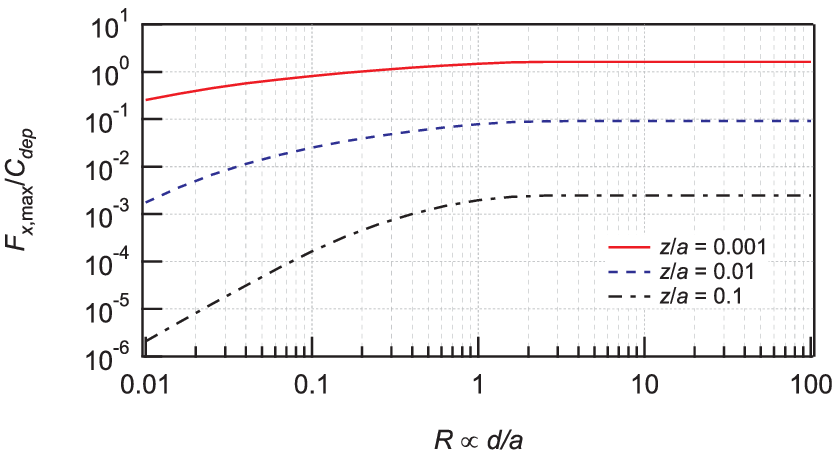}
 }
 \makebox[\textwidth][b]{\hfill {\small a) DEP force maximum value} \hfill}
\end{minipage}
\hfill
\begin{minipage}[t]{0.49\textwidth}
 \makebox[\textwidth][t]{
   \includegraphics[
      width=\textwidth
      ]{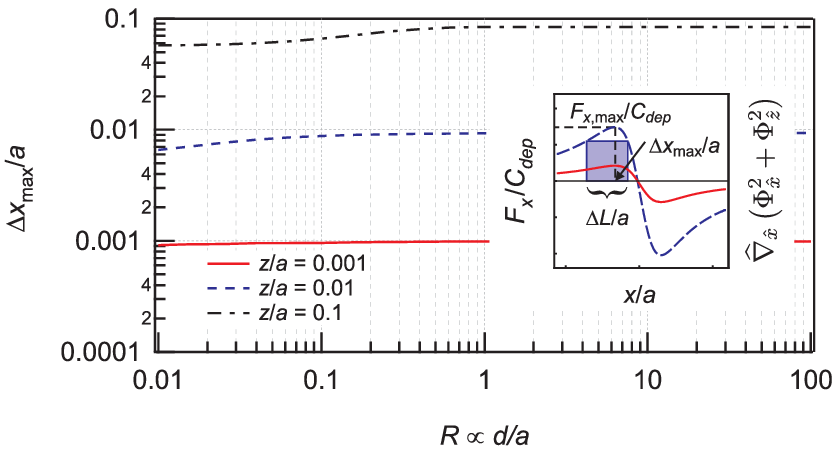}
 }
 \makebox[\textwidth][b]{\hfill {\small  b) DEP force maximum position} \hfill} 
\end{minipage}
\end{minipage}
 \begin{minipage}{\textwidth}
\caption{
Dependence of the $x$-component of the positive DEP force maximum value 
normalized to $C_{dep}$ (a) and the distance $\Delta x_{\rm max}$ of the point,
 where the $x$-component of the DEP force reaches its maximum value (b), from 
the domain wall (normalized to the domain spacing $a$) as a function of the 
parameter $R=(\pi d)/(2a)\sqrt{\varepsilon_a/\varepsilon_c}$. The inset shows 
the maximum value of the DEP force $F_{x,max}/C_{dep}$ at the distance $\Delta 
x_{max}$ from the domain wall and the distance $\Delta L$ over which the DEP 
force has a given minimal value. The parameters $\Delta x_{max}$ and $\Delta 
L$ are measured as a fraction of domain spacing $a$.
\label{fig05}}
 \end{minipage}
\end{center}
\end{figure*}
%
%
%
Both 
Figs.~\ref{fig02} and \ref{fig03} clearly indicate an essential role of the 
geometry of the domain pattern in the numerical value of the dielectrophoretic 
force produced by the bound charges of the surface of the ferroelectric film 
with a domain pattern. To get a clear insight into the role of domain pattern 
geometry on the DEP force, Fig.~\ref{fig05} shows the maximum value of the 
$x$-component of the DEP force and the position of this maximum, which is 
expressed in the distance from the domain wall and measured as a fraction of 
the domain spacing $a$, were calculated as a function of the parameter $R$ for 
three distances $z$ from the surface of the ferroelectric film. 
Figure~\ref{fig05} indicates that there is a qualitative difference in the 
aforementioned dependencies for values of $R$ smaller and greater than one. In 
the systems with a thin ferroelectric film and wide domains, which are 
characterized by $R<1$, the maximum value of the $x$-component of the DEP 
force is decreasing polynomially with a decreasing value of the parameter $R$. 
On the other hand, the systems with a dense ferroelectric domain pattern, which 
is characterized by $R>1$, the maximum of the DEP force is insensitive to the 
value or $R$. It should be noted that the maximum values of the DEP force 
$x$-component at the given distance $z=\hat{z}/k$ from the surface of the 
ferroelectric film are normalized to $C_{dep}$ and they are, therefore, 
dominantly controlled by the geometry of the domain pattern.

\subsection{The role of Brownian motion}
\label{subsec_brownian_motion}

In order to achieve a stable pattern of nanoparticles, the DEP force that acts 
to move the nanoparticle into the trap must overcome the forces that acts on 
that particle due to thermal movements of surrounding medium molecules, i.e. 
the forces that are responsible for the {\em Brownian motion}. The role of 
Brownian motion in DEP trapping of particles has been already analyzed by 
several authors\cite{Washizu.IEEETIndustAppl.30.1994,
Hughes.JPhysDApplPhys.31.1998}. Here we adopt the results achieved by Hughes 
and Morgan\cite{Hughes.JPhysDApplPhys.31.1998} where the limit for particle 
trapping is estimated by comparing the velocities that are induced by DEP 
force with that of the Brownian motion. Result of this analysis yield the 
formula for the estimated critical minimal particle radius that is needed for 
successful trapping of a particle:\cite{Hughes.JPhysDApplPhys.31.1998}
\begin{equation}
	r > r_{crit} = 
	\left(
		\frac{
			10\,k_BT
		}{
			\pi \varepsilon_m \varepsilon_0  K\, \Delta L 
			\left|\nabla E^2\right|
		}
	\right)^{1/3},
	\label{eq19}               
\end{equation}               
where
\begin{equation}
	K =  
	\frac{
		\varepsilon_p - \varepsilon_m
	}{	
		\varepsilon_p + 2\varepsilon_m
	}
	\label{eq20}               
\end{equation}  
is the Clausius-Mossotti factor, $\Delta L$ is typical distance over which the 
field gradient $\nabla E^2$ has a given minimal value (see the inset in 
Fig.~\ref{fig05}b), $k_B$ is the Boltzmann constant and $T$ is the 
thermodynamic temperature. In the notation of this Article, the above formula 
can be rewritten in the form:
\begin{equation}
	r_{crit} = 
	\left(
		\frac{
			10\,k_BT
		}{
			\pi \varepsilon_m \varepsilon_0 K C_E^2\, S(\hat{z}, g, R) 
		}
	\right)^{1/3},
	\label{eq21}               
\end{equation}               
where 
\begin{equation}
	S(\hat{z}, g, R) = 
		\widehat{\Delta L}  
		\left|\widehat{\nabla}_{\hat{x}}
			\left( 
				\Phi_{\hat{x}}^2
				+	
				\Phi_{\hat{z}}^2
			\right)
		\right|_{\rm rms}
	\label{eq22}
\end{equation}
is a dimensionless function given by the product of the effective value of the 
function $|\widehat{\nabla}_{\hat{x}}(\Phi_{\hat{x}}^2 + \Phi_{\hat{z}}^2)|$ 
on the distance $\Delta L$ normalized to the domain spacing $a$, i.e. 
$\widehat{\Delta L}=k\,\Delta L$. For a given material, given values of the 
parameters $g$ and $R$, and a given distance from the surface of the 
ferroelectric film $\hat{z}$, it is possible to find a maximum value of the 
function $S_{\rm max}=\max\{S(\hat{z}, g, R)\}$ by a numerical calculation 
(see Tab.~\ref{tab02}).

\subsection{Conclusions}
\label{subsec_conclusions}

The spatial distribution of the inhomogeneous electric field produced in a 
liquid above the top surface of the polydomain ferroelectric film has been 
calculated. The inhomogeneous electric field is the source of the 
electrophoretic and dielectrophoretic forces that can be used for a deposition 
of nanoparticle pattern. The key element of our analysis, \emtext{considering 
the ionic nature of the liquid}, makes our model quite realistic. Our 
numerical simulations have shown that the intersection of the domain wall with 
the surface of the ferroelectric film represents \emtext{a trap} for 
dielectric nanoparticles in the case of positive dielectrophoresis. The 
analysis of the critical conditions that control the deposition of dielectric 
nanoparticles and the stability of the deposited pattern are presented.

%
%
\begin{table}
\caption{Calculated characteristic numerical parameters of the DEP force 
produced due to spontaneous polarization reversal from a single domain state to 
a polydomain pattern of a given geometry in the selected ferroelectric 
materials. It is considered: $r=1\,{\rm \mu m}$, $z=0.01\,a$, $R=10$, 
$\varepsilon_m=2.66$, $\varepsilon_p=3.0$, and $T=300\,{\rm K}$.
    \label{tab02}}
\begin{ruledtabular}
\begin{tabular}{llcccc}
Parameter& Unit& LiTaO${}_3$& LiNbO${}_3$& BaTiO${}_3$& PbTiO${}_3$ \cr \hline
$g$& 1& 18.1& 18.6& 309.3& 34.0 \cr
$C_E$& GV\,m${}^{-1}$& 30& 41& 14& 41 \cr
$C_{dep}$& ${\rm \mu N}$& 168& 312& 37.5& 321 \cr
$|\nabla E^2|_{\rm max}$& $\rm 10^{18}V^2\,m^{-3}$& 110& 195& 0.09& 60 \cr 
$F_{dep,{\rm max}}/C_{dep}$& $10^{-3}$& 66& 63& 0.25& 20 \cr
$F_{dep,{\rm rms}}$& ${\rm \mu N}$& 8.1& 14.4& 0.007& 4.6 \cr
$S_{\rm max}$& $10^{-3}$& 5.39& 5.14& 0.02& 1.6 \cr
$r_{crit}$& nm& 1.5& 1.2& 15.8& 1.8 
\end{tabular}
\end{ruledtabular}
\end{table}
%
%
%
As an application of our analysis, one can consider two ways of realization of 
DEP force patterning using ferroelectric domains. First, it is the situation 
where the inhomogeneous electric field is produced by a spontaneous 
polarization reversal to form the required domain pattern using advanced 
methods of the nanoscale domain geometry engineering. Considering the values 
of material parameters that correspond to experiments by Grilli {\em et 
al.}\cite{Grilli.ApplPhysLett.92.2008}, i.e. pentanoic acid with 
$\varepsilon_m=2.66$ and flour particles with $\varepsilon_p=3$, the estimated 
numerical parameters of positive DEP force are presented in Tab.~\ref{tab02}. 
It is seen that using this procedure, it is possible to achieve a large local 
values of the DEP forces that results in very small critical radii of 
particles (about 2\,nm) that can be trapped near the domain walls. 
Disadvantage of this method stems from the fact that the bound charges due to 
the discontinuous change of the spontaneous polarization at the surface of the 
ferroelectric film can be partially or completely compensated by free charge 
carriers due to the possible nonzero conductivity of the ferroelectric film or 
due to the ionic nature of the liquid. In this case, the inhomogeneous 
electrostatic field can disappear after some time and other method 
nanoparticle decoration should be used.

%
%
\begin{table}
\caption{Calculated characteristic numerical parameters of the DEP force 
produced due to pyroelectric effect after heating the ferroelectric film from 
the room temperature up by $\Delta T=50$\,K in the selected ferroelectric 
materials.
    \label{tab03}}
\begin{ruledtabular}
\begin{tabular}{llcccc}
Parameter& Unit& LiTaO${}_3$& LiNbO${}_3$& BaTiO${}_3$& PbTiO${}_3$ \cr \hline
$C_E$& GV\,m${}^{-1}$& 0.62& 0.11& 0.54& 1.1 \cr
$C_{dep}$& nN& 73.4& 2.2& 55.5& 222 \cr
$|\nabla E^2|_{\rm max}$& $\rm 10^{15}V^2\,m^{-3}$& 48& 1& 0.1& 40 \cr
$F_{dep,{\rm rms}}$& nN& 3.6& 0.1& 0.01& 3.2 \cr
$r_{crit}$& nm& 20& 64& $140$& 20 
\end{tabular}
\end{ruledtabular}
\end{table}
%
%
%
The second way of the realization of DEP forces using ferroelectric polydomain 
sample is the use of pyroelectric effect. It is known that the pyroelectric 
coefficients in the opposite domains are of the opposite sign. Therefore, when 
the ferroelectric polydomain sample is heated up by $\Delta T$, the bound 
charges of the surface density $\pm\gamma_S\Delta T$ appear on the surface of the 
film due to the pyroelectric effect, where $\gamma_S$ is the pyroelectric 
coefficient and its sign differs from domain to domain. In order to get some 
numerical estimates for this case, the numerical values of parameters $C_E$ 
and $C_{dep}$ should be calculated by replacing the symbol $P_0$ by 
$\gamma_S\Delta T$ in Eqs.~\ref{eq04c} and \ref{eq16b}. Estimated numerical 
parameters of DEP forces achieved by this method are presented in 
Tab.~\ref{tab03}. Advantage of this approach is its relatively easy way of 
realization, e.g. using local laser beam heating of a single domain film or 
uniform heating of a polydomain film. Disadvantage is a presence of smaller 
values of DEP forces that result in greater value of the critical radius of 
particles that can be trapped at required locations (larger than 20\,nm).
The numerical estimates presented in Tab.~\ref{tab03} are in a qualitative 
agreement with experimental results by Grilli {\em et al.} 
\cite{Grilli.ApplPhysLett.92.2008}. 

Finally, we have theoretically demonstrated that using ferroelectric domain 
patterns it is possible to achieve a very flexible way of nanoparticle decoration,
 which has several advantages: (i) using advanced and recently well developed 
methods of domain geometry engineering, it is possible to achieve a 
``rewritable'' systems for the preparation of an arbitrary nanoparticle 
patterns, (ii) since the ferroelectric domain wall is an extremely thin 
spacial feature comparing with other ferroic material or deposited electrodes, 
it is possible to realize patterning of nanoparticles with sub-micron lateral 
dimensions. We believe that the obtained results represent a useful tool that 
can be profitably used for designing the systems with ferroelectric polydomain 
films representing a flexible way for decoration of nanoparticles.
%


\acknowledgments
Authors would like to express their sincere gratitude to Štěpánka Klímková and 
Josef Šedlbauer for many useful discussions concerning the physical chemistry 
aspects of this work. This work has been supported by the Czech Science 
Foundation, Project Nos. GACR~202/06/0411, GACR~202/07/1289, and 
GACR~P204/10/0616. Financial support of the Ministry of Education of the Czech 
Republic (Grant no. MSM0021620835) is gratefully acknowledged.


\bibliographystyle{apsrev}
\bibliography{mokry_cond-mat_v1}

\end{document}